%% file: main_to_arxiv.tex
\documentclass[final,3p,times,twocolumn]{elsarticle}
\usepackage{multirow}
\usepackage{longtable}
\usepackage[hyphens]{url}
\usepackage[]{hyperref}
\hypersetup{breaklinks=true}
\usepackage{comment}
\usepackage{amssymb}
\usepackage{longtable}
\usepackage{booktabs}
\usepackage{qtree}
\usepackage{multirow}
\usepackage{graphicx}
\usepackage[table]{xcolor}
\usepackage{colortbl}
\usepackage{pifont}
\newcommand{\cmark}{\color{black}{\ding{51}}}%
\newcommand{\xmark}{\color{gray}{\ding{55}}}%

\usepackage{tikz,tikz-qtree}

\journal{Computer Speech and Language}

\begin{document}

\begin{frontmatter}


\title{Towards sound based testing of  COVID-19 - Summary of the first Diagnostics of COVID-19 using Acoustics (DiCOVA) Challenge}

\author{Neeraj Kumar Sharma$^*$, Ananya Muguli$^*$, Prashant Krishnan$^*$,  Rohit Kumar,\\ Srikanth Raj Chetupalli, and Sriram Ganapathy\footnote{Corresponding author: Sriram Ganapathy, sriramg@iisc.ac.in \\
$^*$ Equal contribution.}}
\affiliation{organization={Learning and Extraction of Acoustic Patterns (LEAP) Lab, Electrical Engineering,  Indian Institute of Science},
            city={Bangalore},
           country={India}}

\begin{abstract}
The technology development for point-of-care tests (POCTs) targeting respiratory diseases has witnessed a growing demand in the recent past. Investigating the presence of acoustic biomarkers in modalities such as cough, breathing and speech sounds, and using them for building POCTs can offer fast, contactless and inexpensive testing. In view of this, over the past year, we launched the ``Coswara'' project to collect cough, breathing and speech sound recordings via worldwide crowdsourcing. With this data, a call for development of diagnostic tools was announced in the Interspeech 2021 as a special session titled ``Diagnostics of COVID-19 using Acoustics (DiCOVA) Challenge''. The goal was to bring together researchers and practitioners interested in developing acoustics-based COVID-19 POCTs by enabling them to work on the same set of development and test datasets. As part of the challenge, datasets with breathing, cough, and speech sound samples from COVID-19 and non-COVID-19 individuals were released to the participants. The challenge consisted of two tracks. The Track-1 focused only on cough sounds, and participants competed in a leaderboard setting. In Track-2, breathing and speech samples were provided for the participants, without a competitive leaderboard. The challenge attracted $85$ plus registrations with $29$ final submissions for Track-1. This paper describes the challenge (datasets, tasks, baseline system), and presents a  focused summary of the various systems submitted by the participating teams. An analysis of the results from the top four teams showed that a fusion of the scores from these teams yields an area-under-the-curve of $95.1$\% on the blind test data. By summarizing the lessons learned, we foresee the challenge overview in this paper to help accelerate  technology for acoustic-based POCTs.
\end{abstract}




\end{frontmatter}


\section{Introduction}
\label{sec:intro}
\noindent The viral respiratory infection caused by the novel coronavirus, SARS-CoV-2, termed as the coronavirus disease 2019 (COVID-19), was declared a pandemic by the World Health Organization (WHO) in March 2020. The current understanding of COVID-19 prognosis suggests that the virus infects the nasopharynx and then spreads to the lower respiratory tract~\cite{schaefer2020situ}. One of the key strategies to combat the rapid spread of infection across populations is to perform rapid and large-scale testing.

The prominent COVID-19 testing methodologies currently take a molecular sensing  approach. The current gold-standard technique, termed as reverse transcription polymerase chain reaction (RT-PCR)~\cite{corman2020detection}, relies on using nasopharyngeal or throat swab samples. The swab sample is treated with chemical reagents enabling isolation of the ribonucleic acid (RNA), followed by deoxyribonucleic acid (DNA) formation, amplification, and analysis, facilitating the detection of COVID-19 genome in the sample. However, this approach has several limitations. The swab sample collection procedure violates physical distancing~\cite{who_poct}. The processing of these samples requires a well equipped laboratory, with readily available chemical reagents and expert analysts. Further, the turnaround time for test results can vary from several hours to a few days. The protein based rapid antigen testing (RAT)~\cite{peeling2021scaling} improves over the speed of detection while being inferior to the RT-PCR in performance. The RAT test also involves the need of chemical agents.

\input{table_litReview}
 
In view of the above mentioned limitations in RT-PCR/RAT testing, there is a need to design highly specific, rapid and easy-to-use point-of-care tests (POCTs) that could identify the infected individuals in a decentralized manner. Using acoustics for developing such a POCT would overcome  various limitations in terms of speed, cost and scalability, and allows remote testing.

\subsection{Exploring acoustics based testing}
\noindent Acoustics for diagnosis of  pertussis \cite{pramono2016cough}, tuberculosis \cite{botha2018detection}, childhood pneumonia \cite{abeyratne2013cough}, and asthma \cite{hee2019development} were explored using cough sounds recorded with portable devices. As COVID-19 is an infection affecting the respiratory pathways~\cite{li2021epidemiology}, recently, researchers have made efforts towards acoustic data collection. A list of acoustic datasets is provided in Table \ref{table:literature_review}. Building on these datasets, few studies have also evaluated the possibility of COVID-19 detection using acoustics. Brown et al.~\cite{brown2020exploring} used cough and breathing sounds jointly and attempted a binary classification task of separating COVID-19 infected individuals from healthy. The dataset was collected through crowd-sourcing, and the analysis was done on $141$ COVID-19 infected individuals. The authors reported a performance between $80-82\%$ AUC (area-under-curve). Agbley et al. \cite{agbley2020wavelet} demonstrated $81\%$ specificity (at $43\%$ sensitivity) on a subset of the COUGHVID dataset \cite{orlandic2020coughvid}. Imran et al.~\cite{imran2020ai4covid} studied cough sound samples from four groups of individuals, namely, healthy, and those with bronchitis, pertussis, and COVID-19 infection. They report an accuracy of $92.64\%$. Laguarte et al. \cite{laguarta} used a large sample set of COVID-19 infected individuals and report an AUC performance of $90\%$. Andreu-Perez et al. \cite{9361107} created a more controlled dataset via collecting cough sound samples from patients visiting hospitals, and they report $99\%$ AUC on a COVID-19 pool of $2339$ coughs.

\begin{figure*}[t]
    \centering
    \includegraphics[width=14cm, height=4cm]{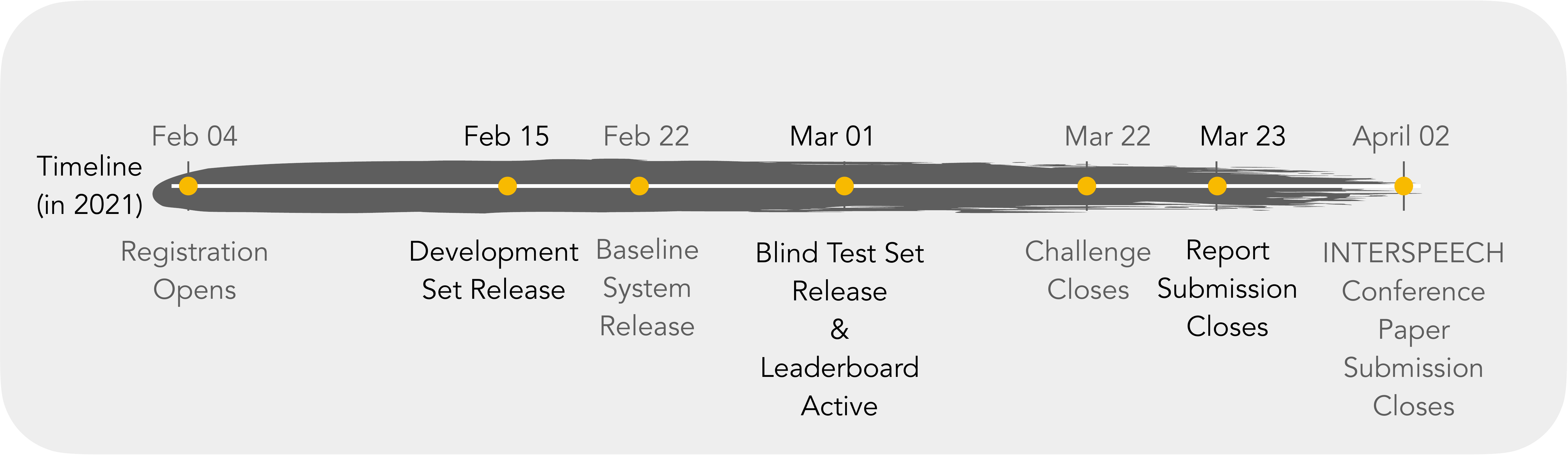}
    \caption{The DiCOVA challenge timeline.}
    \label{fig:chall_participation}
\end{figure*}


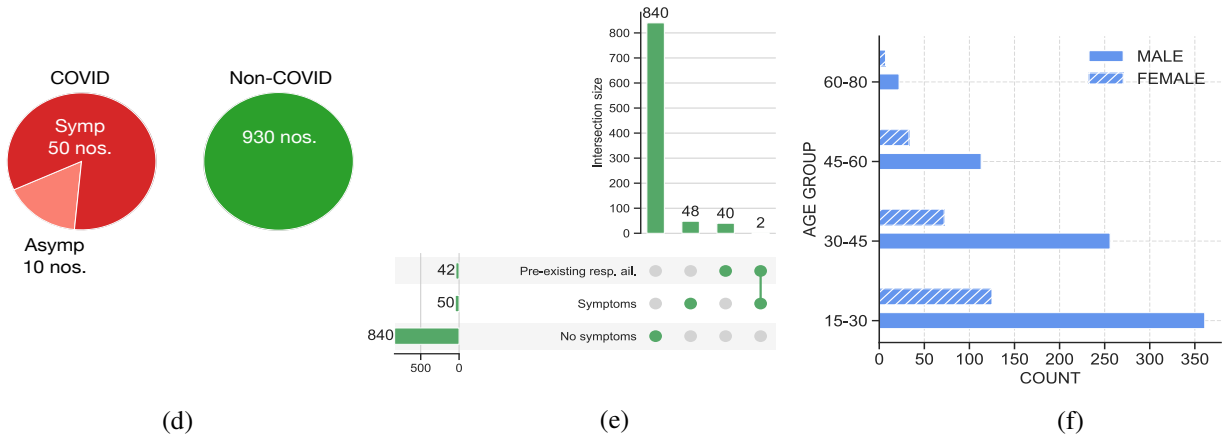
\begin{figure*}[t]
    \centering
    \input{plot_metadata}

    \caption{An illustration of Track-1 and Track-2 development datasets. Here, (a,c) show the COVID and non-COVID pool size in terms of number of individuals; (b,d) show the breakdown of non-COVID individuals into categories of no symptoms, symptoms (cold, cough), and pre-existing respiratory ailment (asthma, chronic lung disease); (c,d) depicts the age group distribution in the development dataset.}
    \label{fig:chall_dataset}
\end{figure*}

Although these studies are encouraging, they suffer from some challenging limitations. Some of these studies are based on privately collected, small datasets. Further, the ratio of COVID-19 patients to healthy (or non-COVID) is different in every study. The performance metrics are also different across studies. Some of the studies report performance per-cough bout, and others report per-patient.  Further, most of the studies have not bench-marked on other open source datasets, making it difficult to compare among the various propositions.  

\subsection{Contribution}
\noindent We launched the ''Diagnostics of COVID-19 using Acoustics (DiCOVA) Challenge''~\cite{dicova} with two primary goals. Firstly, to encourage the speech and audio researchers to analyze acoustics of cough and speech sounds for a problem of immediate societal relevance. The challenge was launched under the umbrella of Interspeech 2021, and participants were given an option to submit their findings to a special session in this flagship conference. Secondly, and more importantly, to provide a benchmark for monitoring the progress in acoustic based diagnostics of COVID-19. The development and (blind) test datasets were provided to the participants to facilitate design of classifier systems. A leaderboard was created allowing participants to rank order their performance against others. This paper describes the details of the challenge including the dataset, the baseline system (Section~\ref{sec:2}), and provides a summary of the various submitted systems (Section~\ref{sec:3}). An analysis of the scores submitted by the top teams (Section~\ref{sec:4}), and the insights gained from the challenge are also presented (Section~\ref{sec:5}). 

\section{DiCOVA Challenge}
\label{sec:2}
\noindent The DiCOVA challenge\footnote{\url{https://dicova2021.github.io/}} was launched on $4-$Feb, 2021 and the challenge lasted till $23-$Mar, $2021$. A timeline of the challenge is shown in Figure~\ref{fig:chall_participation}. 
The participation was through a registration process followed by the release of development and test datasets. A remote server based scoring system was created with a leaderboard setting. This provided near real-time ranking of teams, and monitoring progress on the blind test set\footnote{\url{https://competitions.codalab.org/competitions/29640}}. The call for participation in the challenge attracted $85$ plus registrations. Further, $29$ teams made final submissions on the blind test set.

\subsection{Dataset}
\noindent The challenge dataset is derived from the Coswara dataset \cite{sharma2020coswara}, a crowd-sourced dataset of sound recordings. The Coswara data is collected using a website\footnote{\url{https://coswara.iisc.ac.in/}}. The volunteers of all age groups and health conditions were requested to record their sound data in a quiet environment using a mobile web connected device. 

The participants initially  provide demographic information like age and gender. An account of their current health status in form of a questionnaire of symptoms as well as pre-existing conditions like respiratory ailments and co-morbidity are recorded. The web based tool also records the result of the COVID test conducted and the possibility of exposure to the virus through primary contacts.  

The acoustic data from each subject contains $9$ audio categories, namely, $(a)$ shallow and deep breathing ($2~$nos.), $(b)$ shallow and heavy cough ($2$~nos.), $(c)$ sustained phonation of vowels [\ae] (as in bat), [i] (as in beet), and [u] (as in boot) ($3$~nos.), and $(d)$ fast and normal pace number counting ($2$~nos.).  The dataset collection protocol was approved by the Human Ethics Committee of the Indian Institute of Science, Bangalore, and P. D. Hinduja National Hospital and Medical Research Cente, Mumbai, India. 

The DiCOVA Challenge used a subset of the Coswara dataset, sampled from the data collected between April-$2020$ and Feb-$2021$. The sampling  included only age group of $15-80$ years. The subjects with health status of ``recovered'' (who were COVID positive however fully recovered from the infection) and ``exposed'' (suspecting exposure to the virus) were not included in the dataset.  Further, subjects with audio recordings of duration less than $500$~msec were discarded. The resulting curated subject pool was divided into the following two groups.
\begin{itemize}
\item \textbf{non-COVID}: Subjects self reported as healthy, having symptoms such as cold/cough/fever or  having pre-existing respiratory ailments (like asthma, pneumonia, chronic lung disease) but were not tested positive for COVID-19.
\item \textbf{COVID}: Subjects self-declared as COVID-19 positive (symptomatic with mild/moderate infection or asymptomatic)
\end{itemize} 

\noindent The DiCOVA 2021 challenge featured two tracks. The Track-1 dataset composed of (heavy) cough sound recordings from $1040$ subjects. The Track-2 dataset is composed of deep breathing, vowel [i], and number counting (normal pace) speech recordings from $992$ subjects. For each track, the dataset was divided into training with validation (development set)  and test set.

An illustration of the important metadata details in the development set is provided in Figure~\ref{fig:chall_dataset}. About $70$\% of the subjects were male. The majority of the participants lie in the age group of $15-40$ years. Also, the dataset is highly imbalanced with less than $10$\% of the participants belonging to the COVID category. We retained this class imbalance in the challenge as this addresses the typical real-world POCT scenario.


\begin{figure*}[t]
    \centering
    \input{plots_datasplits}
    \caption{Illustration of dataset splits for Track-1 (cough) and Track-2 (breathing and speech).}
    \label{fig:my_label}
\end{figure*}
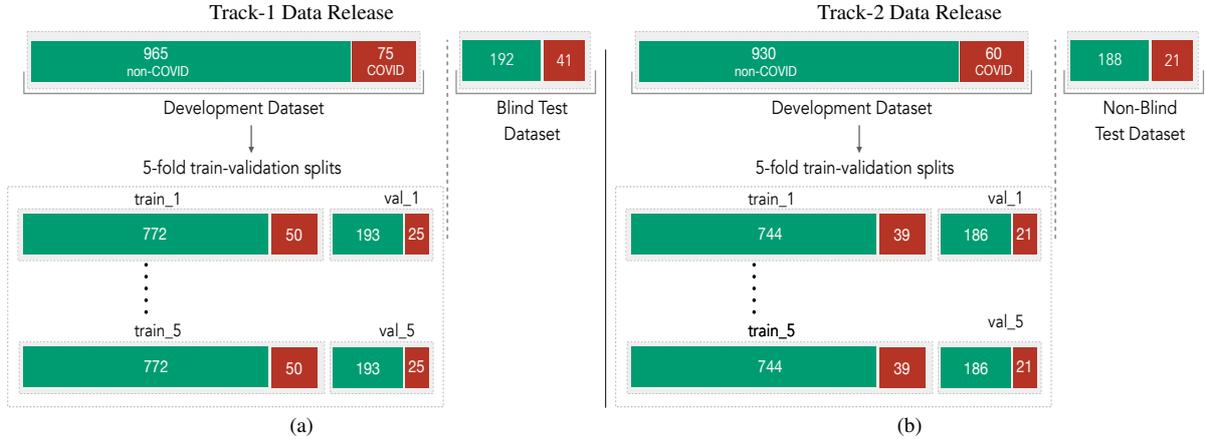

\subsection{Audio Specifications}
\noindent The crowd-sourced dataset collection acts as a good representation of real-world data with sensor variability arising from diverse recording devices. For the challenge, we re-sampled all audio recordings to $44.1$~kHz and compressed them to FLAC (Free Lossless Audio Codec) format for ease of distribution. The average duration of Track-1 development set cough recordings is $4.72~($standard error $(S.E.) \pm 0.07)$~sec. The average duration of Track-2 development set audio recordings is -breathing $17.72~(\pm 0.68)$~sec, vowel [i] $12.40~(\pm 0.17)$~sec, and number counting speech $14.71~(\pm 0.11)$~sec.

\subsection{Task}
\begin{itemize}
    \item\textbf{Track 1}:
    The task was based on cough samples.  This was the primary track of the challenge with most teams participating only in this track. We released the baseline system for this track as well.  The development dataset release contained a five-fold validation setup. A leaderboard website  was hosted for the challenge enabling teams to evaluate their system performance (validation and blind test). The participating teams were required to submit the COVID probability score for each audio file in the validation and test sets. The tool computes the area-under-curve (AUC) and specificity/sensitivity. Every team was provided a maximum of $25$ tickets for evaluation over the course of the challenge.
    
    \item\textbf{Track 2}:
    Track-2 explored the use of recordings other than cough for the task of COVID diagnostics.  The audio recordings released in this track composed of breathing, speech related to sustained phonation of vowel [i] and number counting ($1-20$). The development and (non-blind) test sets were released concurrently, without any formal leaderboard style evaluation and competition.
\end{itemize}
The data and the baseline system setup were provided to the registered teams after signing the terms and conditions document. As per the document, the teams were not allowed to use the publicly available Coswara dataset\footnote{\url{https://github.com/iiscleap/Coswara-Data}}.  

\subsection{Evaluation Metrics}
\noindent The focus of the task (Track-1 and Track-2) was binary classification. As the dataset was imbalanced, we choose not to use accuracy as an evaluation metric. Each team submitted COVID probability scores ($\in [0,1]$ with higher values indicating more likelihood of COVID infection) for the list of validation/test audio recordings. For performance evaluation, we used the scores with the ground truth labels to compute the receiver operating characteristics (ROC) curve. The curve was obtained by varying the decision threshold between $0-1$ with a step size of $0.0001$. The area under the resulting ROC curve was used as a performance measure for the classifier, where the area was computed using the trapezoidal method. The AUC formed the primary evaluation metric. Further, specificity (true negative rate), at a sensitivity (true positive rate) greater than or equal to $80\%$ was also used as secondary evaluation metric.

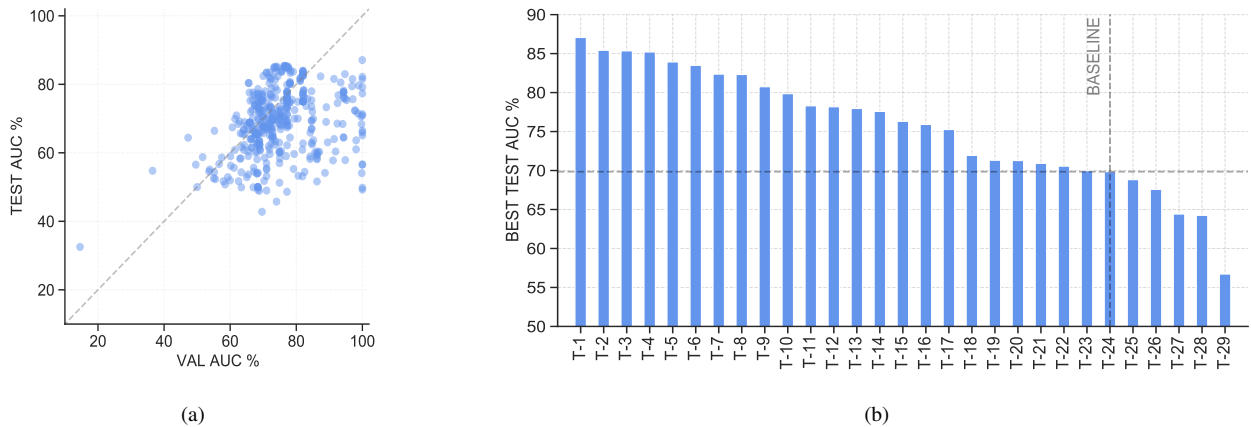
\begin{figure*}[t!]
    \centering
    \input{plots_val_test_aucs}
    \vspace{-0.8cm}
    \caption{(a) A scatter plot of the average five-fold validation AUC versus test AUC performance for every submission on the leaderboard. (b) Test set AUC performance in rank ordered manner for each of the system submissions.}
    \label{fig:summary_val_test_aucs}
\end{figure*}

\subsection{Baseline System}
\noindent The baseline system was implemented using the  \texttt{scikit-learn} Python library~\cite{scikit-learn}.

\noindent \textit{Pre-processing}: For every audio file, the signal was normalized in amplitude. Using a sound activity detection threshold of $0.01$ and a buffer size of $50~$msec on either side of a sample, any region of the audio signal with amplitude lower than threshold was discarded. Also, initial and final $20$~ms snippets of the audio were removed to avoid abrupt start and end activity in the recordings.

\begin{table}[t!]
\centering
\begin{tabular}{@{}cllrr@{}}
\toprule
\textbf{Track} & \textbf{Sound} & \multicolumn{1}{c}{\textbf{Model}} & \multicolumn{2}{c}{\textbf{\begin{tabular}[c]{@{}c@{}}Performance \\ (AUC\%)\end{tabular}}} \\ \cmidrule(l){4-5} 
 &  &  & Val. (std. err) & Test \\ \midrule
\textbf{} & \textbf{} & LR & 66.95 ($\pm$ 1.74) & 61.97 \\
1 & Cough & MLP & 68.80 ($\pm$ 1.05) & \textbf{69.85} \\
 &  & RF & 70.69 ($\pm$ 1.39) & 67.59 \\ \midrule
\textbf{} & \textbf{} & LR & 60.95 ($\pm$ 2.17) & 60.94 \\
 & Breathing & MLP & 72.47 ($\pm$ 1.96) & 71.52 \\
 &  & RF & 75.17 ($\pm$ 1.23) & \textbf{76.85} \\ \cmidrule(l){2-5} 
 & \textbf{} & LR & 71.48 ($\pm$ 0.55) & 67.71 \\
2 & Vowel {[}i{]} & MLP & 70.39 ($\pm$ 1.84) & 73.19 \\
 &  & RF & 69.73 ($\pm$ 1.93) & \textbf{75.47} \\ \cmidrule(l){2-5} 
 & \textbf{} & LR & 68.93 ($\pm$ 1.09) & 61.22 \\
 & Speech & MLP & 73.57 ($\pm$ 0.71) & 61.13 \\
 &  & RF & 69.61 ($\pm$ 1.56) & \textbf{65.27} \\ \bottomrule
\end{tabular}
\caption{The baseline system performance on Track-1 and Track-2 on the development set (5-fold val) and test set.}
\label{tab:baseline_aucs}
\end{table}

\noindent \textit{Feature extraction}: The baseline system used the $13$ dimensional mel-frequency cepstral coefficients (MFCCs), its delta and delta-delta coefficients, computed over $1024$ samples ($23.2$ms), with a hop of $441$ samples ($10$ms). The resulting feature dimension was $39\times1$.  \\
\noindent \textit{Classifiers}: The following three classifiers were designed.
\begin{itemize}
    \item Logistic Regression (LR): The LR clasifier was trained for $25$ epochs. The binary cross entropy (BCE) loss with a $l_2$   regularization strength of $0.01$ was used for optimizing the model. 
    \item Multi layer perceptron (MLP): A single-layer perceptron model with $25$ hidden units and \texttt{tanh()} activation was used. Similar to the LR model, the BCE loss with a $\ell_2$ regularization of strength 0.001 was optimized for parameter estimation. The loss was optimized using Adam optimizer with an initial learning rate of $0.001$. The COVID samples were over-sampled to compensate for the data imbalance (weighted BCE loss). 
    \item Random Forests (RF): A random forest classifier was trained with $50$ trees using \textit{Gini} impurity criterion for tree growing. 
\end{itemize}
\noindent \textit{Inference and performance}:
To obtain a classification score for an audio recording: $(i)$ the file is pre-processed, $(ii)$ frame-level MFCC features are extracted, $(iii)$ frame-level probability scores are computed using the trained model(s), and $(iv)$ all the frame scores are averaged to obtain a single COVID probability score for the audio recording. For evaluation on the test set files, the probability scores from five validation models (for a classifier) are averaged to obtain the final score.

The performance of the three classifiers used in the baseline system is provided in Table \ref{tab:baseline_aucs}. All classifiers performed better than chance. For Track-1, the AUC for the test set was better for the MLP classifier ($69.85\%$ AUC). For Track-2, RF gave the best AUC in all sound categories ($65.27\%-76.85\%$ AUC).

Further, among the category of acoustic sounds, the breathing samples provided the best AUC ($76.85\%$) performance followed by vowel sound [i] ($75.47\%$). The baseline system code\footnote{\url{https://github.com/dicova2021/Track-1-baseline}} was provided to the participants as a reference for setting up a processing, and scoring pipeline.

\begin{table*}[t!]
\centering
\rowcolors{3}{white}{gray!10}
\begin{tabular}{llclllllclrlr}
\hline
\multirow{2}{*}{\begin{tabular}[c]{@{}l@{}}Team ID\\ rank-wise\end{tabular}} &  & \multicolumn{7}{c}{Implementation} &  & \multicolumn{3}{c}{Performance} \\ \cline{3-9} \cline{11-13} 
 &  & \begin{tabular}[c]{@{}c@{}}Data\\ aug.\end{tabular} &  & \multicolumn{1}{c}{Features} &  & \multicolumn{1}{c}{Classifiers} &  & \multicolumn{1}{l}{Ensemble} &  & \begin{tabular}[c]{@{}r@{}}Test\\ AUC\%\end{tabular} &  & \begin{tabular}[c]{@{}r@{}}Test\\ Spec.\%\end{tabular} \\ \cline{1-1} \cline{3-3} \cline{5-5} \cline{7-7} \cline{9-9} \cline{11-11} \cline{13-13} 
T-1~\cite{mahanta_dicova} &  & \cmark &  & MFCCs &  & CNN &  & \xmark &  & 87.07 &  & 83.33 \\
T-2~\cite{chang_dicova} &  & \cmark &  & mel-spectrogram &  & ResNet50 &  & \cmark &  & 85.43 &  & 82.29 \\
T-3~\cite{harvill_dicova}$\dagger$ &  & \cmark &  & mel-spectrogram &  & LSTM &  & \cmark &  & 85.35 &  & 71.88 \\
T-4~\cite{sodergren_dicova}$\dagger$ &  & \xmark &  & openSMILE &  & RF, SVM &  & \cmark &  & 85.21 &  & 81.25 \\
T-5~\cite{singh_dicova} &  & \cmark &  & MFCC &  & \begin{tabular}[c]{@{}l@{}}CNN, LSTM\\ResNet \end{tabular} &  & \cmark &  & 83.93 &  & 70.83 \\
T-6~\cite{das_dicova}$\dagger$ &  & \xmark &  & ERB-spectrogram &  & RF, MLP &  & \cmark &  & 83.49 &  & 77.08 \\
T-7~\cite{elizalde2021covid19} &  & \xmark &  & YAMNet,OpenL3 &  & Extra Trees &  & \xmark &  & 82.37 &  & 72.92 \\
T-8$^{*}$ &  & \xmark &  & mel-spectrogram &  & ResNet34 &  & \xmark &  & 82.32 &  & 67.19 \\
T-9~\cite{flavio_dicova}$\dagger$ &  & \cmark &  & openSMILE &  & SVM, CNN &  & \cmark &  & 80.75 &  & 63.54 \\
T-10$^{*}$ &  & \cmark &  & mel-spectrogram &  & ResNet34 &  & \cmark &  & 79.86 &  & 68.23 \\
T-11$^{*}$ &  & \cmark &  & mel-spectrogram &  & VGG13 &  & \cmark &  & 78.30 &  & 50.52 \\
T-12~\cite{karas_dicova}$\dagger$ &  & \xmark &  & \shortstack{openSMILE, \\embeddings} & & SVM,LSTM & & \cmark & & 78.18 & & 58.85 \\
T-13$^{*}$ &  & \xmark &  & \begin{tabular}[c]{@{}l@{}}openSMILE, \\mel-spectrogram \end{tabular}&  &
\begin{tabular}[c]{@{}l@{}}DNN, VGG\\CNN\end{tabular} &  & \cmark &  & 77.96 &  & 59.38 \\
T-14$^{*}$ &  & \xmark &  & \begin{tabular}[c]{@{}l@{}}handcrafted, \\log mel-spectrogram \end{tabular} &  & MLP &  & \xmark &  & 77.58 &  & 60.42 \\
T-15~\cite{kamble2021panacea}$\dagger$ &  & \xmark &  & TECC+$\Delta+\Delta\Delta$ &  & LightGBM &  & \xmark &  & 76.31 &  & 53.65 \\
T-16~\cite{banerjee2021residual} &  & \cmark &  & mel-Spectrograms &  & ResNet50 &  & \xmark &  & 75.91 &  & 62.50 \\
T-17$^{*}$ &  & \xmark &  & MFCCs &  & CNN &  & \xmark &  & 75.26 &  & 55.21 \\
T-18~\cite{swapnil_dicova}$\dagger$ &  & \xmark &  & MFCCs &  & DNN &  & \xmark &  & 71.94 &  & 47.40 \\
T-21~\cite{adria_dicova}$\dagger$ &  & \xmark &  & mel-Spectrogram Images &  & ResNet18 &  & \xmark &  & 70.91 &  & 47.40 \\
T-24\cite{dicova}$\dagger$ &  & \xmark &  & MFCCs &  & MLP &  & \xmark &  & 69.85 &  & 53.65 \\
T-27~\cite{gauri_dicova}$\dagger$ &  & \xmark &  & handcrafted features &  & LSTM &  & \xmark &  & 64.42 &  & 40.10 \\
\hline
\end{tabular}
\caption{Summary of submitted systems in terms of feature and model configurations. The specificity(\%) is reported at a sensitivity of $80\%$. Here, $\dagger$: denotes report accepted in Interspeech 2021, $^{*}$ denotes team did not give consent for public release of report.}
\label{table:submitted_systems_review}
\end{table*}

\section{Overview of Submitted Systems}
\label{sec:3}
\noindent A total of $28$ teams participated in the Track-1 leaderboard. Out of these, $20$ teams submitted their system reports describing the explored approaches\footnote{The system reports are available at \url{https://dicova2021.github.io/\#reports}.}. In this section, we provide a brief overview of the submissions, emphasizing on the feature extraction approaches, data augmentation methods, classifier types, and model performances. The Track-2 submission did not require system report submission. Hence, we limit the discussion of the system highlights to Track-1 submissions only.

\subsection{Performance Summary}
\noindent The performance summary of all the submitted systems on the validation and the blind test data is given in Figure~\ref{fig:summary_val_test_aucs}. Figure~\ref{fig:summary_val_test_aucs}(a) depicts a comparison of the validation and test results. Interestingly, there is a slight positive correlation between test and validation performance. For some teams, the validation performances exceed $90$\% AUC. Deducing from the system reports, these outliers are primarily due to training/over-fitting to the validation data. Figure~\ref{fig:summary_val_test_aucs}(b) depicts the best AUC posted by $29$ participating teams (including baseline) on the blind test data. The best AUC performance on the test data was $87.07\%$ AUC, a significant improvement over the baseline AUC ($69.85\%$). 

In total, $23$ out of the $28$ teams reported a performance better than the baseline system. We refer to the teams with Team IDs corresponding to their rank on the leaderboard, that is, best AUC performance as T-1 and so on.

\begin{figure*}[t]
    \centering
    \input{plots_top_4_auc_bars}
    \caption{Illustration of AUC performance on full test set, and test set split by gender, and age. For male set: $171$ nos. ($27$ COVID), for female  set: $62$ nos. ($14$ COVID), for age$<40$ set: $180$ nos. ($31$ COVID), and for age$\geq40$ set: $53$ nos. ($10$ COVID).}
    \label{fig:top_3_auc_gen_age}
\end{figure*}
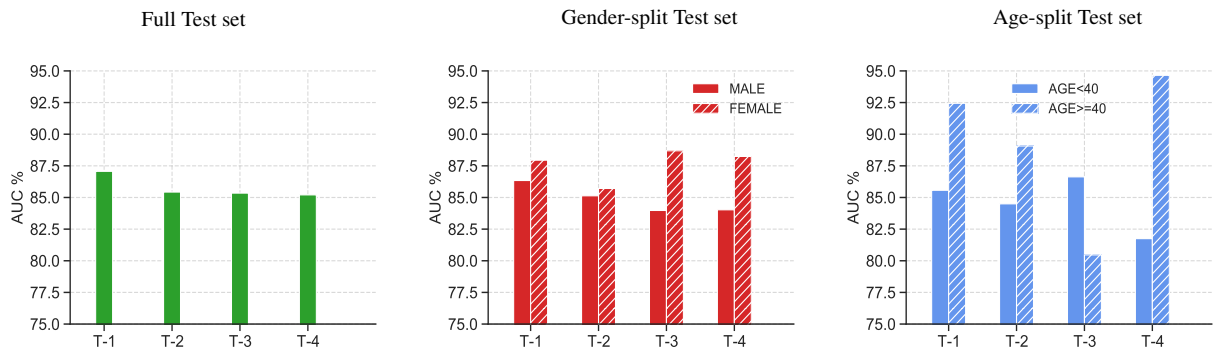
\subsection{Feature and classifier configurations}
\noindent The teams designed and experimented with a wide spectrum of features and classifiers. A concise highlight is shown in Table~\ref{table:submitted_systems_review}. We elaborate on this below.

\subsubsection{Features explored}
\noindent A majority of the teams used mel-spectrograms, mel-frequency cepstral coefficients \cite{davis_mfcc}, or equivalent rectangular bandwidth (ERB) \cite{smith1999bark} spectrograms ($15$ submissions out $21$). Further, the openSMILE features~\cite{opensmile}, which consist of statistical measures extracted on low-level acoustic feature descriptors, were explored by $4$ teams. Few teams explored features derived using Teager energy based cepstral coefficients (TECC \cite{kamble2019analysis}; T-15), and pool of short-term features such as short-term energy, zero-crossing rate, and voicing (T-5, T-14, T-27).
Other teams resorted to using embeddings derived from pre-trained neural networks as features. These included VGGish \cite{vggish}, DeepSpectrum~\cite{Amiriparian2017}, OpenL3~\cite{OpenL3}, YAMNet \cite{yamnet} embeddings (T-7, T-12), and x-vectors \cite{xvectors} (T-15).

\subsection{Classifier models}
\noindent The teams explored various classifier models. These included classical machine learning models, such as decision trees, random forests (RFs), and support vector machines (SVMs), and modern deep learning models, such as convolutional neural networks (CNNs), long short-term memory (LSTM) networks, and residual networks (ResNet). Several teams also attempted an ensemble of models to improve the final system performance. 

The CNNs were explored by teams (T-1, T-5, T-10, T-17). Variants of CNNs with residual connections and recording level average pooling to deal with the variable length input were developed by teams (T-2, T-5, T-8, T-9, T-10, T-13, T-16, T-21). Citing the improved ability of LSTMs to handle variable length inputs, (T-3, T-5, T-12, T-27) explored these models. The classical ML approaches of random forest, logistic regression and SVMs were used by (T-4, T-6, T-12, T-18). LightGBM (Gradient Boosting Machine)~\cite{lightgbm} model was explored by (T-15), and extra trees classifiers were studied by (T-7). 
Pre-training was also studied in several systems (T-3, T-17). Autoencoder style pre-training was used by (T-17). Several teams had also experimented with transfer learning from architectures pre-trained on image based models (T-2, T-8, T-13) and audio based models (T-10, T-13).

\subsection{Model Ensembling}
\noindent The fusion of scores from different classifier architectures was explored by multiple teams (T-3, T-4,T-6 T-10, T-11, T-12, T-13). The fusion of multiple features was explored by (T-13). Further, (T-2, T-3) investigated score fusion of outputs obtained from the model tuned on the five validation folds.

\subsection{Data Augmentation}
\noindent Data augmentation is a popular strategy in which external or synthetic audio data is used in training of deep network models. Few teams ($5$ nos.) reported using this strategy by including COUGHVID cough dataset \cite{orlandic2020coughvid} (publicly available), adding Gaussian noise at varying SNRs, or doing audio manipulations (pitch shifting, time-scaling, etc., via tools such as audiomentations\footnote{\url{https://github.com/iver56/audiomentations}}). Few teams also used data augmentation approaches to circumvent the problem of class imbalance. These included T-1 using mixup~\cite{zhang2017mixup}, (T-3,T-9,T-11) using SpecAugment~\cite{park2019specaugment}, (T-2, T-5, T-9) using additive noise, T-21 using sample replication, and T-5 using Vocal-Tract Length Perturbation (VTLP)~\cite{Jaitly_vocaltract}, to increase the sample counts of the minority class.

Besides these, other strategies for training included gender aware training (T-21), using focal loss~\cite{focal_loss} objective function (T-2,T-8,T-11), and 
hyper-parameter tuning using model searching algorithm TPOT (T-7) \cite{tpot}.

In the next section, we discuss in detail the approaches used by the four top performing teams.

\section{Top performers}
\label{sec:4}
\input{table_auc_sensitivity}
\subsection{T-1: The Brogrammers}
\noindent The team~\cite{mahanta_dicova} focused on using a multi-layered CNN network architecture. Special emphasis was laid on having a small number of learnable parameters. Every audio segment was trimmed or zero padded to $7$~secs. For feature extraction, this segment was represented using $15$ dimensional MFCC features per frame, and a matrix of $15\times 302$ frames was obtained. A cascade of a CNN and fully connected layers, with max-pooling and ReLU non-linearities, was used in the neural network architecture. For data augmentation, the team used the audiomentations tool. The classifier was trained using binary cross entropy (BCE) loss to output COVID probability score. The team did not report performing any system combination unlike several other participating teams. 
\subsection{T-2: NUS-Mornin system}
\noindent The team focused~\cite{chang_dicova} on using  the residual network (ResNet) model with spectrogram images as features. To overcome the limitations of data scarcity and imbalance, the team resorted to three key strategies. Firstly, data augmentation was done by adding Gaussian noise to spectrograms. Secondly, focal loss function was used instead of cross entropy loss. Thirdly, the ResNet50 was pre-trained on ImageNet followed by fine-tuning on DiCOVA development set and an ensemble of four models was used to generate final COVID probability scores.

\subsection{T-3: UIUC SST system}
\noindent The team~\cite{harvill_dicova}  used long short term memory (LSTM) models. With the motivation of generative modeling of mel-spectrogram for capturing informative features of cough, the team proposed  using the auto-regressive predictive coding (APC)~\cite{oord2018representation}. The APC is used to pre-train the initial LSTM layers operating on the input mel-spectrogram. The additional layers of the full network, which was  composed of BLSTM and fully connected layers, was trained using the DiCOVA  development set. As the number of model parameters was high, the team also used data augmentation using COUGHVID dataset~\cite{orlandic2020coughvid} and SpecAugment~\cite{park2019specaugment} tool. The binary cross entropy was chosen as the loss function. 
The final COVID-19 probability score was obtained as an average of several similar models, trained on development data subsets or sampled at different checkpoints during training.

\subsection{T-4: The North System}
\noindent The team~\cite{sodergren_dicova}  explored classical machine learning models like random forests (RF), support vector machines (SVM), and multi-layer perceptron (MLP) rather than deep learning models. The features used were the $6373$ dimensional openSMILE functional features~\cite{opensmile}. The openSMILE features were z-score normalized to prevent feature domination. The hyper-parameters of the models were tuned to obtain the best results. The SVM models alone provided an AUC of $85.1$\% on the test data. The RF and the MLP scored an AUC of $82.15$ and $75.65$, respectively. The final scores were obtained by a weighted average of the probability scores from the RF and SVM models, with weights of $0.25$ and $0.75$, respectively.

\subsection{Top 4 teams: Fairness}
\noindent Here, we present a fairness  analysis of the scores generated by the top 4 teams. We particularly focus on gender-wise and age-wise performance on the test set. Figure~\ref{fig:top_3_auc_gen_age} depicts this performance. Interestingly, all the four teams gave a better performance for female subjects. Similarly, the test dataset was divided into two groups based on subjects with age$<40$ and age$\geq40$. Here, the top two teams had a considerably higher AUC for age$\geq40$ subjects, while T-3 had a lower AUC for this age group and T-4 had the highest. In summary, the performance of top four teams did not reflect the bias in the development data ($70\%$ male participants, largely in age $\leq 40$ group).

\subsection{Top 4 teams: Score Fusion}
\begin{figure}[t!]
    \centering
    \includegraphics[width=8cm, height=8cm]{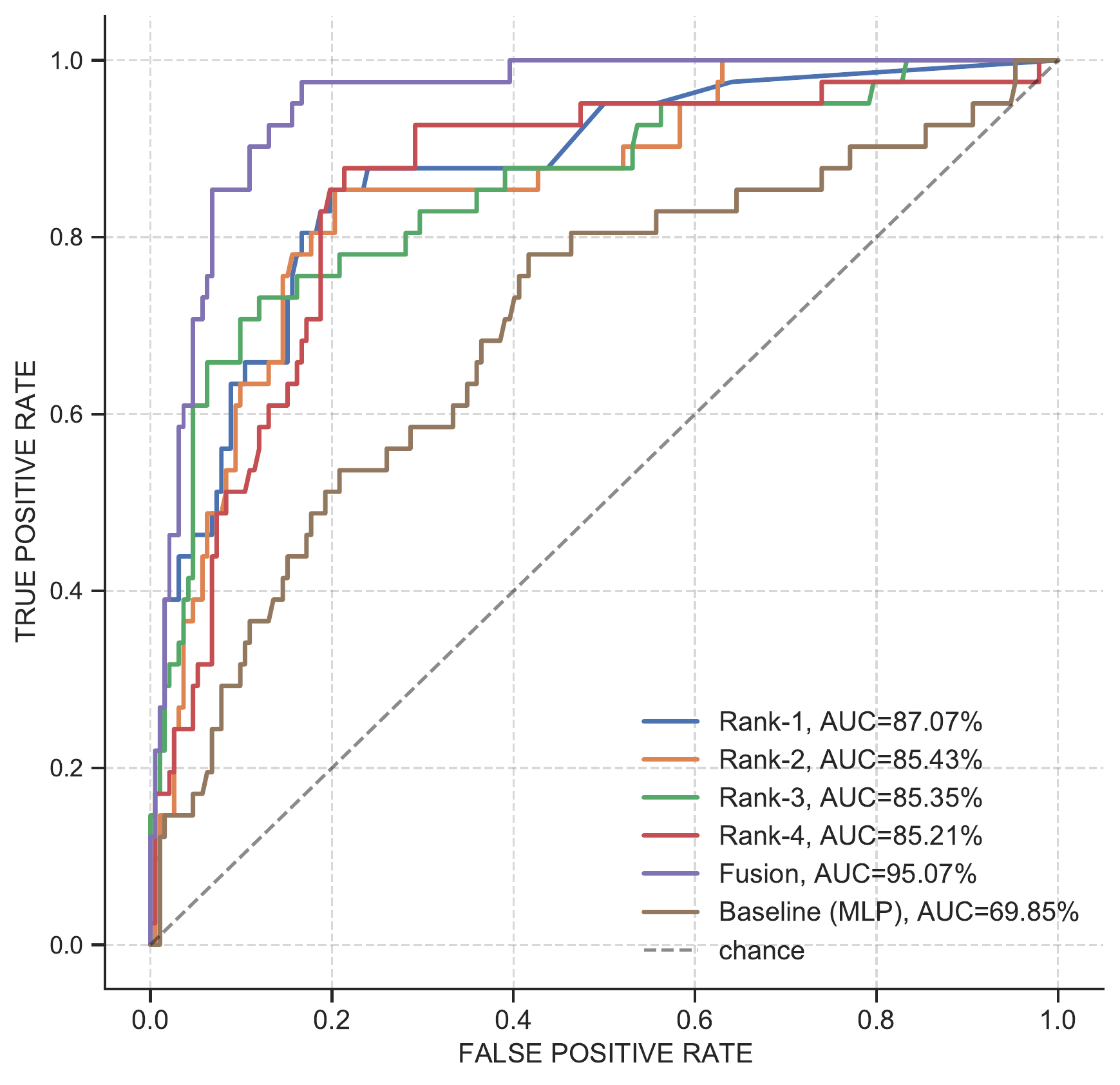}
    \caption{Illustration of ROCs obtained on the test set for the top four teams. The ROC associated with the hypothetical score fusion system obtained using the top four teams is also shown.}
    \label{fig:score_fusion}
\end{figure}

The systems from the top four  teams differ in terms of features, model architectures, and data augmentation strategies.   We consider a simple arithmetic mean fusion of the scores from top $4$ teams. Let $p_{ij}$,~$1 \leq i \leq N$ and $1 \leq j \leq T$ be the COVID probability score predicted by the $j^{th}$ team submission for the $i^{th}$ subject in the test data. Here, $N$ denotes the number of subjects in the test set, and  $T$ is four. The scores are first calibrated by correcting for the range as
\begin{equation}
    \hat{p}_{ij} = \frac{p_{ij} - p_{min,j}}{ p_{max,j}-p_{min,j} },
\end{equation}
where $p_{min,j} = \min\left( p_{1j},\dots,p_{Nj} \right)$ and $p_{max,j} = \max\left( p_{1j},\dots,p_{Nj} \right)$.  The fused scores are obtained as,
\begin{equation}
    p_{i,f} = \frac{1}{4} \sum_{j=1}^4 \hat{p}_{ij}.
\end{equation}
\noindent The ROC obtained using these prediction scores is denoted by \textit{Fusion} in Figure~\ref{fig:score_fusion}. This gives an AUC of $95.10\%$, a significant improvement over each of the individual system results. Table~\ref{table:auc_sensitivity} depicts the sensitivity of the top four systems, the fusion, and baseline (MLP) at $95$\% specificity. The fused model surpasses all the other models and achieves a sensitivity of $70.7$\%. 

\section{Discussion}
\label{sec:5}
\subsection{Challenge accomplishments}
\noindent The challenge problem statement for Track-1 required the design of a binary classifier. A clear problem statement, with a well-defined evaluation metric (AUC), encouraged a significant number of registrations. This included $85$ plus teams from around the globe. We also noticed a good representation from both industry and academia. The $28$ teams which completed the challenge came from $9$ different countries. Additionally, $8$ teams associated themselves with industry. Among the submissions, $23$ out of the $28$ teams exhibited a performance well above the baseline system AUC (see Figure~\ref{fig:summary_val_test_aucs}(b)). 

Altogether, the challenge provided a common platform for interested researchers to explore a timely diagnostic problem of immense societal impact. The results indicate potential in using acoustics for COVID-19 POCT development. The challenge turnaround time was $49$~days, and the progress made by different teams in this short time span highlighted the efforts that were undertaken by the community.  

Several works in this challenge will be presented at the  DiCOVA Special Session, Interspeech 2021 (to be held during $30$~Aug-$3$ Sept, 2021). Following the peer review process, the special session will feature $11$ accepted papers.

The World Health Organization (WHO) has stated that a sensitivity of $\geq70\%$ (at a specificity of $95\%$) is necessary for an acceptable POCT tool \cite{who_poct}. The top four teams fell short of this benchmark (see Table~\ref{table:auc_sensitivity}), indicating that there is scope for further development in future. However, a simple combination of the scores from the systems of these teams achieves this benchmark. This suggests ways to reap advantage via collaboration between multiple teams for improved tool development. The development of such a sound based diagnostic tool for COVID-19 diagnosis would offer multiple advantages in terms of speed, cost and remote testing. 

\subsection{Limitations and Future Scope}
\noindent  The challenge, being first of its kind, also had its own limitations. The dataset provided was largely imbalanced, with a majority of the samples belonging to the non-COVID class. Although the imbalance reflects the prevalence of the infection in the population, it will be ideal to improve this imbalance in future challenges. The Coswara dataset \cite{sharma2020coswara} is being regularly updated, and as of June 2021, it contains data from approximately $200$ COVID-19 positive individuals and $2000$ non-COVID individuals. However, at the time of the challenge, the size of the COVID positive class was only about $120$ in the dataset. 

A majority of the DiCOVA dataset samples came from India. While the cultural dependence of cough and breathing is not well established, it will be ideal to evaluate the performance on datasets collected from multiple geographical sites. Towards this, future challenges can include demographically balanced datasets, with close collaborations between multiple sites involved in the data collection efforts.

The task involved in the challenge simplified to a binary classification setting. However, in a practical scenario, there are multiple respiratory ailments resulting from bacterial, fungal, or viral infections, with each condition potentially leaving a unique biomarker. The future evaluation of respiratory ailments may target multi-class categorization, which will also widen the usability of the tool.

The data did not contain information regarding the progression of the disease (or the time elapsed since the positive COVID-19 test). Also, the participants in the ``recovered'' and ``exposed'' category were not analyzed in the challenge. The leaderboard and system highlights reported were limited to the cough recordings only. As seen in Table~\ref{tab:baseline_aucs}, analysis using breathing and speech signals can also yield performance results comparable to those observed in cough recordings. In addition, the Coswara tool~\cite{sharma2020coswara} also records the symptom data from participants.  The combination of all acoustic categories with symptoms in developing the tool might further push the performance metrics of these tools to surpass the regulatory requirements. 

In the DiCOVA challenge, the performance ranking of the teams was based on AUC metric, which just conveyed the significance of the model's ability to perform binary classification. However, the challenge did not emphasize model interpretability and explainability as key requirements. In a healthcare scenario, the interpretability of the model decisions may be as important as the accuracy. Hence, future challenges should encourage this aspect. Additionally in future, it's important to focus on reproducibility of the models as well as lower memory/computational foot-print, which will benefit the rapid development of a tool.

\section{Acknowledgments}
\noindent The authors would like to thank the Department of Science and Technology (DST), Government of India, for providing financial support to the Coswara Project through the RAKSHAK programme. The authors would like to thank the Organizing Committee, Interspeech 2021 for giving us the opportunity to host this challenge under the umbrella of ISCA. The authors would  like to express their gratitude to Anand Mohan for the design of the web based data collection platform, and Dr. Nirmala R., Dr. Shrirama Bhat,  Dr. Lancelot Pinto, and Dr. Viral Nanda for their coordination in data collection.


\bibliographystyle{elsarticle-num} 
\bibliography{mybib}

\end{document}

%% file: table_litReview.tex
\begin{table*}[t]
\centering
\rowcolors{2}{white}{gray!10}
\begin{tabular}{lllllllllll}
\hline
\bf{Ref.} &  & \bf{Name} &  & \bf{Category} &  & \bf{Access} &  & \begin{tabular}[c]{@{}l@{}}\# \bf{of COVID}/ \\ \bf{non-COVID}  \end{tabular} &  & \bf{Method} \\ \cline{1-1} \cline{3-3} \cline{5-5} \cline{7-7} \cline{9-9} \cline{11-11} 
Orlandic et. al.~\cite{orlandic2020coughvid} &  & Coughvid &  & Cough &  & Public &  & 1010/19062 &  & Crowdsource\\
Sharma et. al.~\cite{sharma2020coswara} &  & Coswara &  & \begin{tabular}[c]{@{}l@{}}Breathing, Cough,\\ Vowels, Speech,\\ Symptoms\end{tabular} &  & Public &  & 153/1402 &  & Crowdsource\\
Wilson et. al.~\cite{virufy_set} &  & Virufy &  & Cough &  & Public &  & 7/9 &  & Hospital \\
Cohen et. al.~\cite{nocoda} &  & NoCoCoDa &  & Cough &  & On request &  & 13/NA &  & YouTube\\
Brown et. al.~\cite{brown2020exploring} &  & \begin{tabular}[c]{@{}l@{}}COVID-19\\ Sounds\end{tabular} &  & Cough, Breathing &  & On request &  & 235/6378 &  & Crowdsource\\
\hline
\end{tabular}
\caption{A list of COVID-19 acoustics accessible datasets.}
\label{table:literature_review}
\end{table*}

%% file: plot_metadata.tex
\begin{tikzpicture}
\node[inner sep=0pt,anchor=center] (main_fig) at (8,12) 
{\includegraphics[width=16cm,height=5cm]{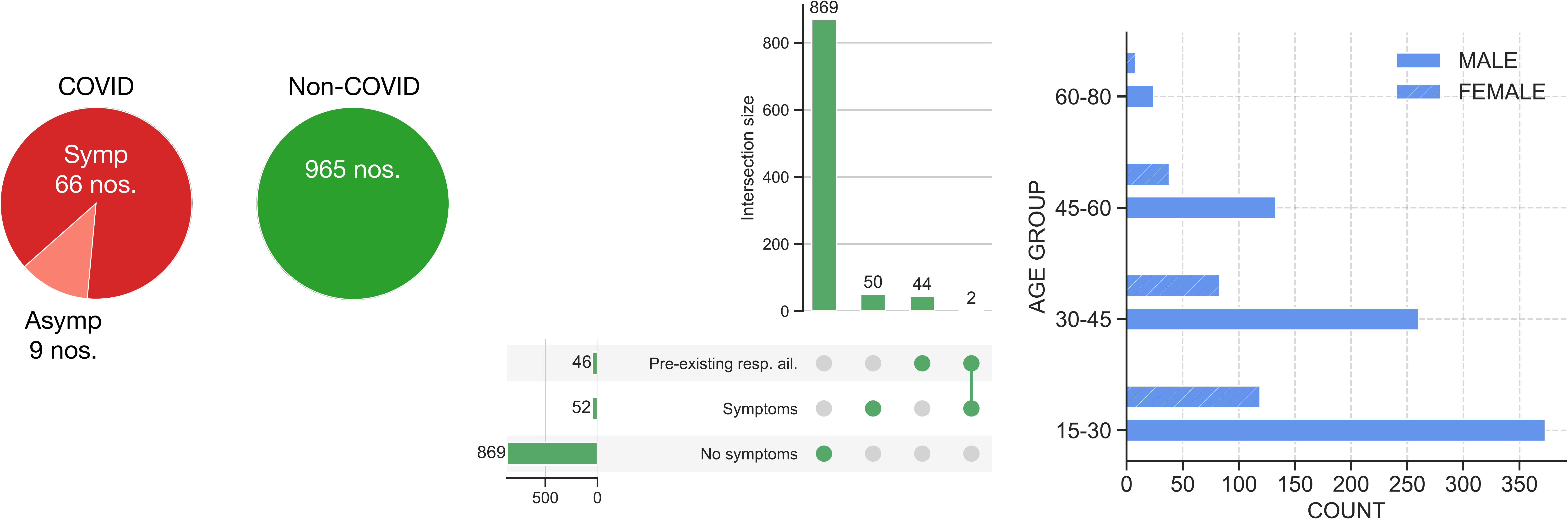}};

\node[inner sep=0pt,anchor=center] (main_fig) at (8,5) 
{\includegraphics[width=16cm,height=5cm]{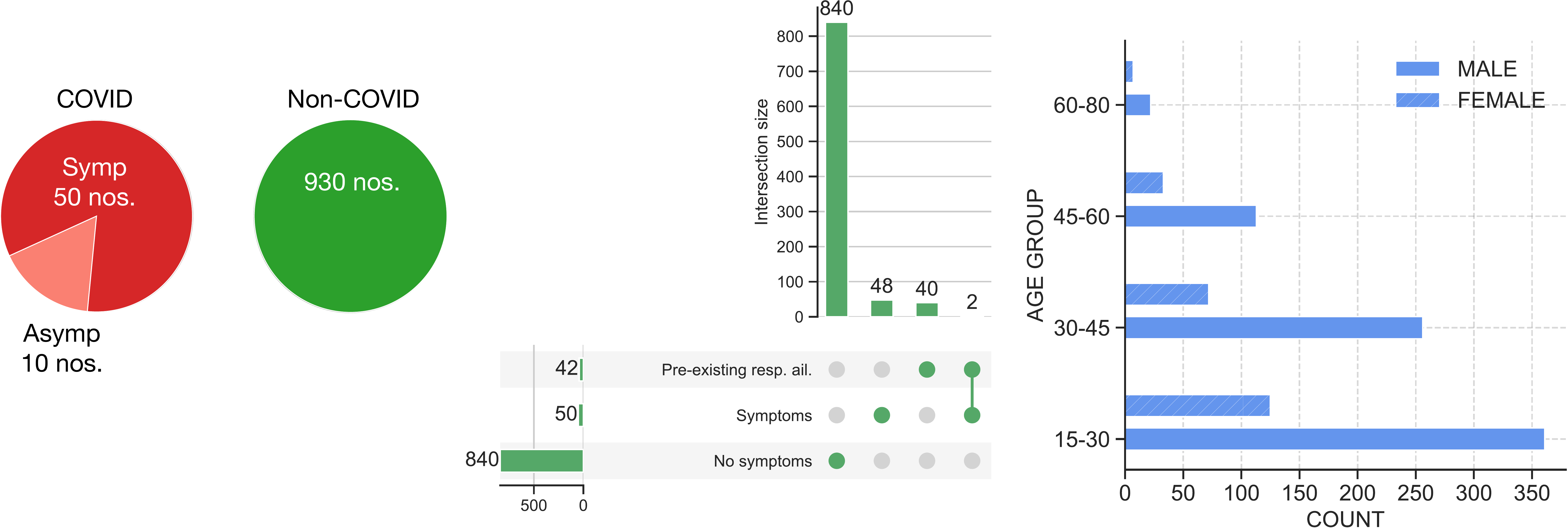}};

\node[font=\fontsize{10}{8}\selectfont,rotate=0,anchor=center] at (8,15) (l1) {Track-1 Development Dataset Metadata};
\node[font=\fontsize{10}{8}\selectfont,rotate=0,anchor=center] at (2.25,9) (l1) {(a)};
\node[font=\fontsize{10}{8}\selectfont,rotate=0,anchor=center] at (8,9) (l1) {(b)};
\node[font=\fontsize{10}{8}\selectfont,rotate=0,anchor=center] at (14,9) (l1) {(c)};

\draw[line width=.1 mm, dashed] (2,8.5) -- (14,8.5);

\node[font=\fontsize{10}{8}\selectfont,rotate=0,anchor=center] at (8,8) (l1) {Track-2 Development Dataset Metadata};
\node[font=\fontsize{10}{8}\selectfont,rotate=0,anchor=center] at (2.25,2) (l1) {(d)};
\node[font=\fontsize{10}{8}\selectfont,rotate=0,anchor=center] at (8,2) (l1) {(e)};
\node[font=\fontsize{10}{8}\selectfont,rotate=0,anchor=center] at (14,2) (l1) {(f)};

\end{tikzpicture}

%% file: plots_datasplits.tex
\begin{tikzpicture}
\node[inner sep=0pt,anchor=center] (main_fig) at (2,10) 
{\includegraphics[width=7.75cm,height=5cm]{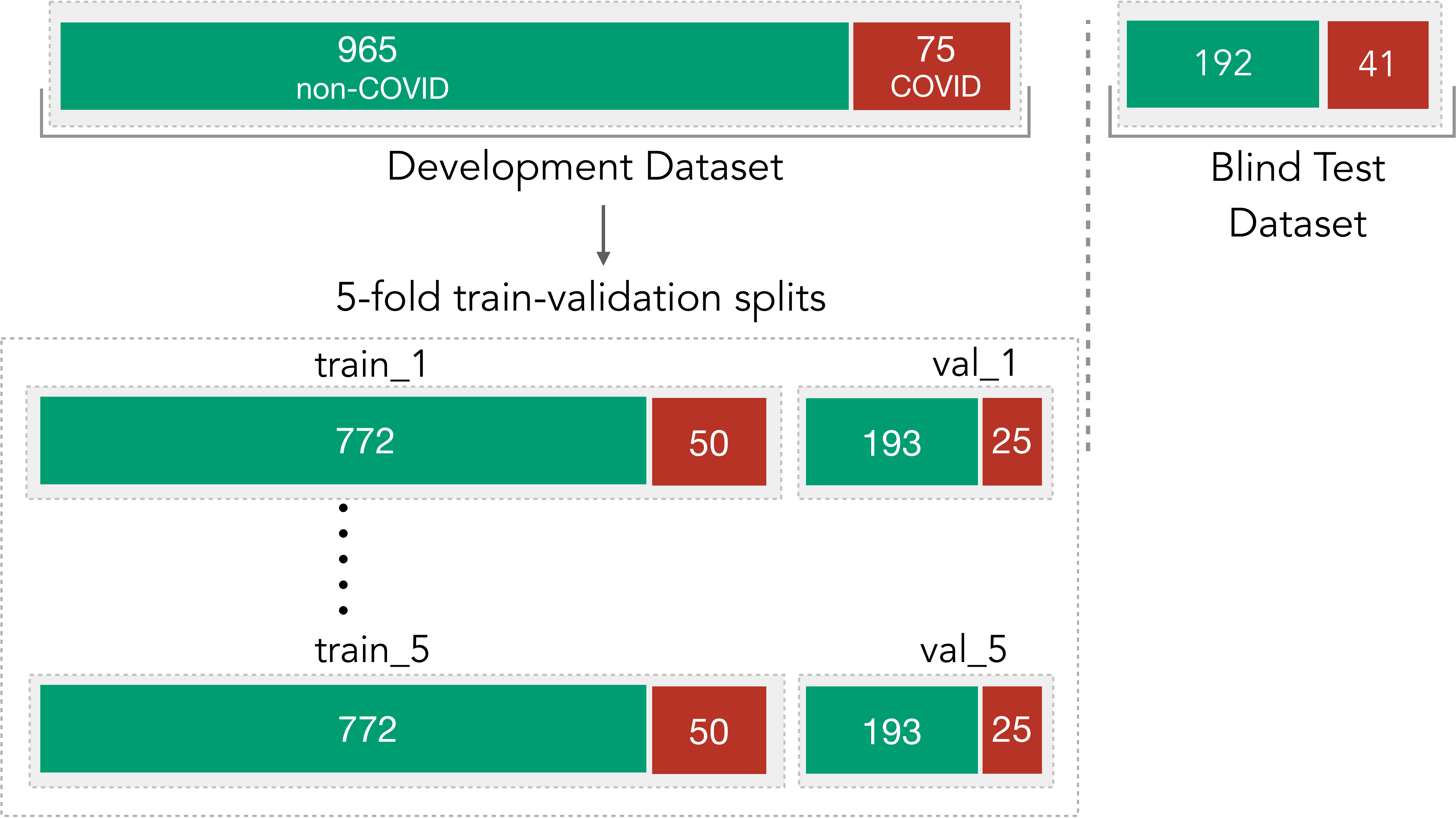}};

\node[inner sep=0pt,anchor=center] (main_fig) at (10,10) 
{\includegraphics[width=7.75cm,height=5cm]{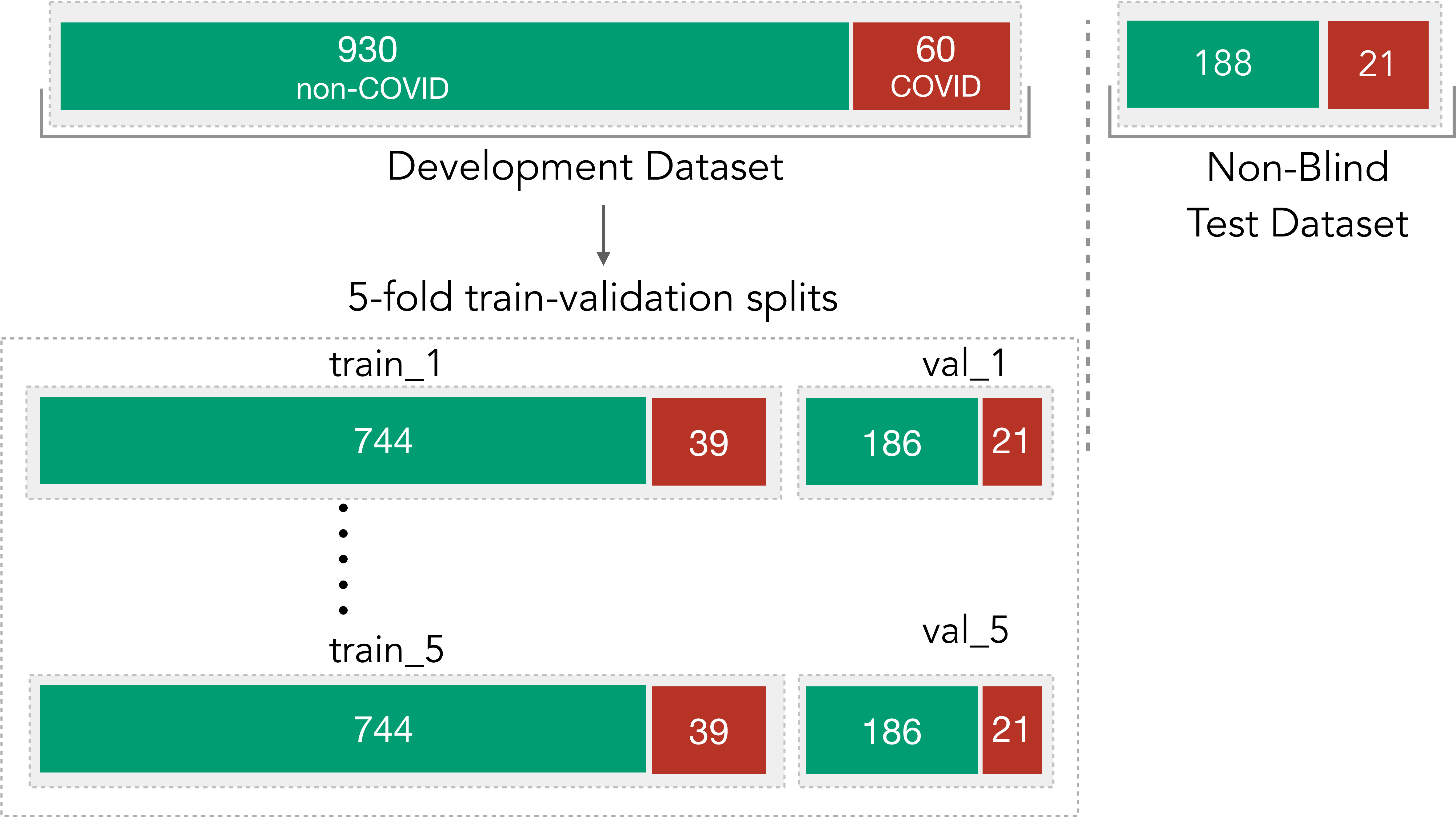}};

\node[font=\fontsize{8}{6}\selectfont,rotate=0,anchor=center] at (2,12.75) (l1) {Track-1 Data Release};
\node[font=\fontsize{8}{6}\selectfont,rotate=0,anchor=center] at (10,12.75) (l1) {Track-2 Data Release};

\node[font=\fontsize{8}{6}\selectfont,rotate=0,anchor=center] at (2,7.25) (l1) {(a)};
\node[font=\fontsize{8}{6}\selectfont,rotate=0,anchor=center] at (10,7.25) (l1) {(b)};

\draw[line width=.1 mm] (6,7.5) -- (6,12.25);

\end{tikzpicture}

%% file: plots_val_test_aucs.tex
\begin{tikzpicture}
\node[inner sep=0pt,anchor=center] (main_fig) at (2,10) 
{\includegraphics[width=5cm,height=5cm]{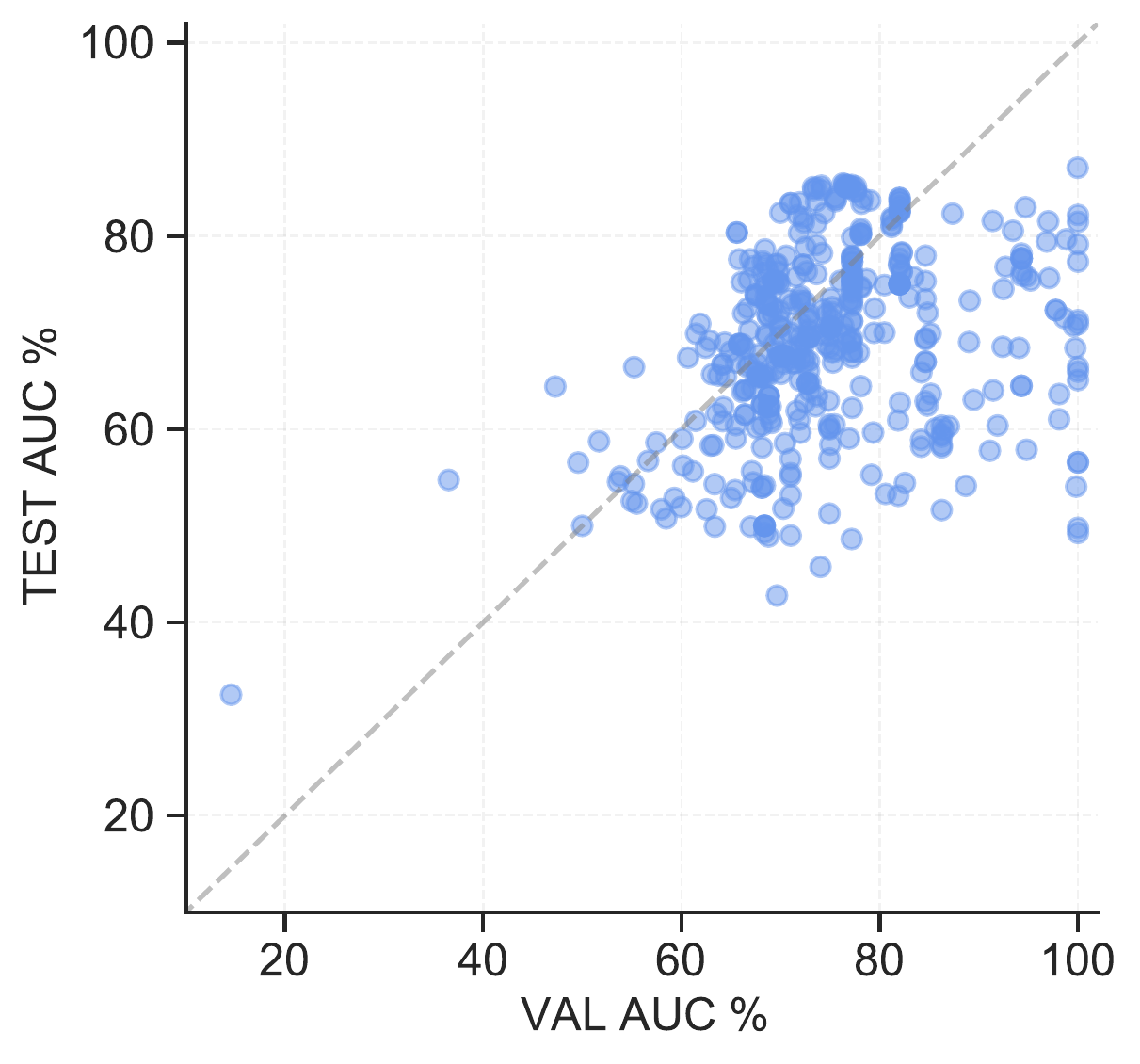}};

\node[inner sep=0pt,anchor=center] (main_fig) at (11,10) 
{\includegraphics[width=10cm,height=5cm]{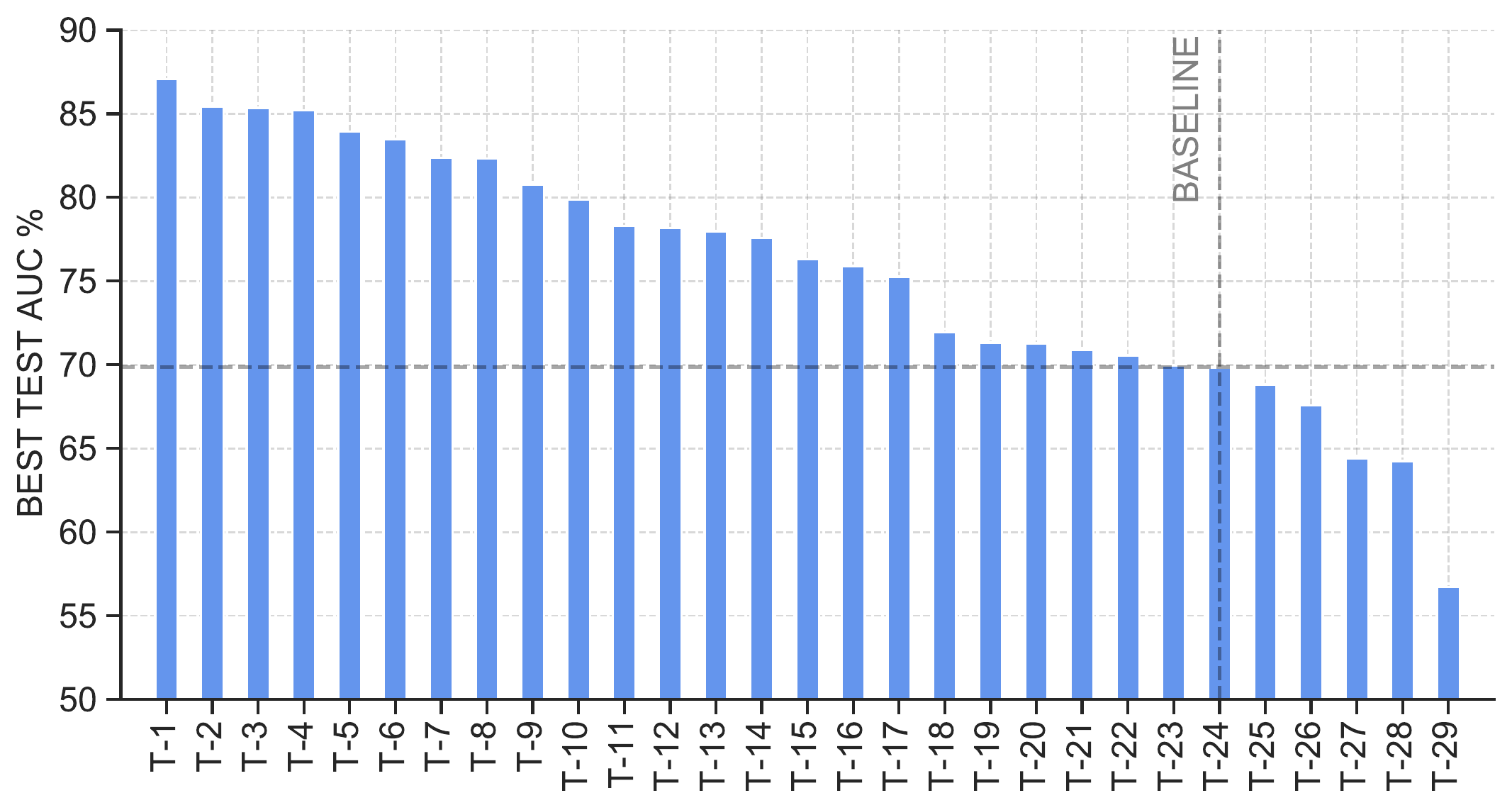}};

\node[font=\fontsize{8}{6}\selectfont,rotate=0,anchor=center] at (2,7) (l1) {(a)};
\node[font=\fontsize{8}{6}\selectfont,rotate=0,anchor=center] at (11,7) (l1) {(b)};


\end{tikzpicture}

%% file: plots_top_4_auc_bars.tex
\begin{tikzpicture}
\node[inner sep=0pt,anchor=center] (main_fig) at (2,10) 
{\includegraphics[width=5cm,height=4cm]{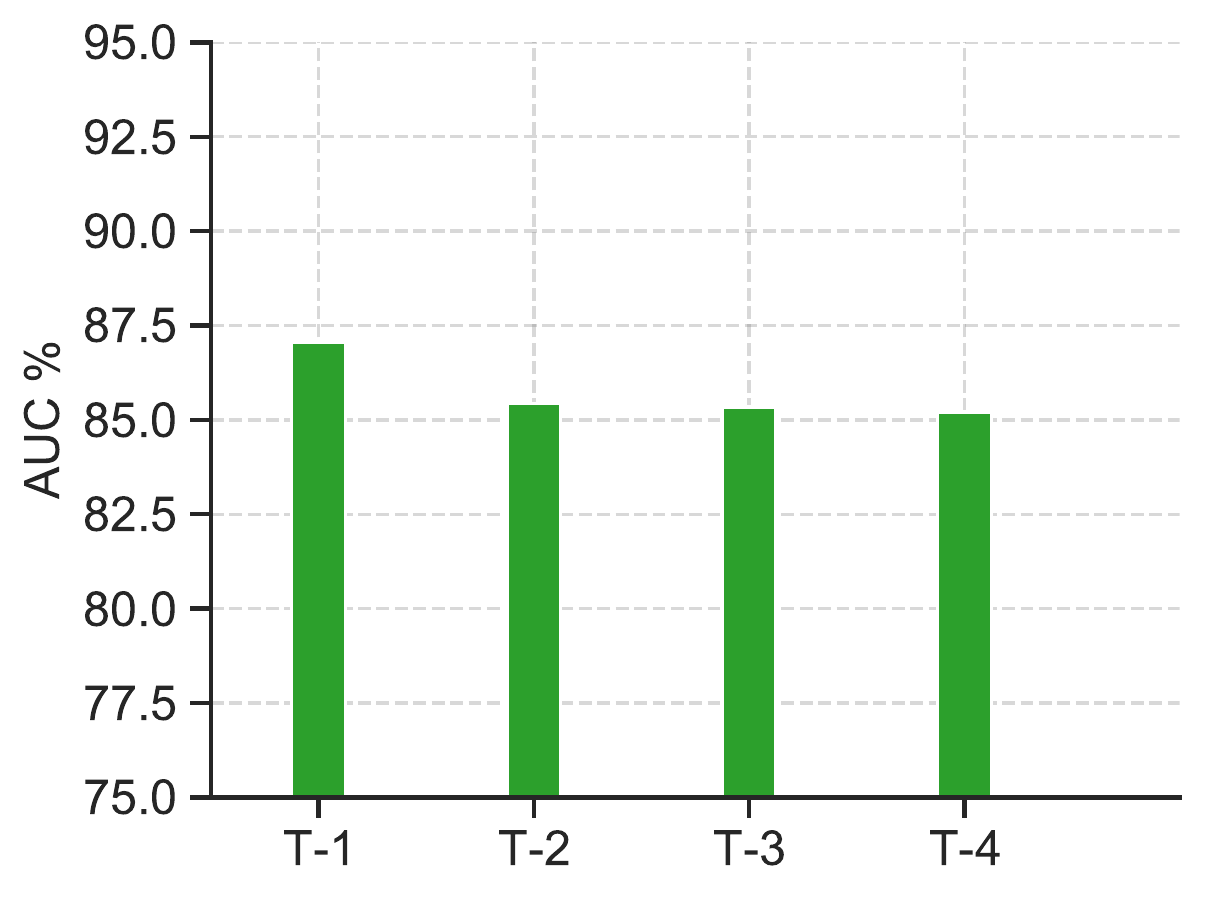}};

\node[inner sep=0pt,anchor=center] (main_fig) at (7.5,10) 
{\includegraphics[width=5cm,height=4cm]{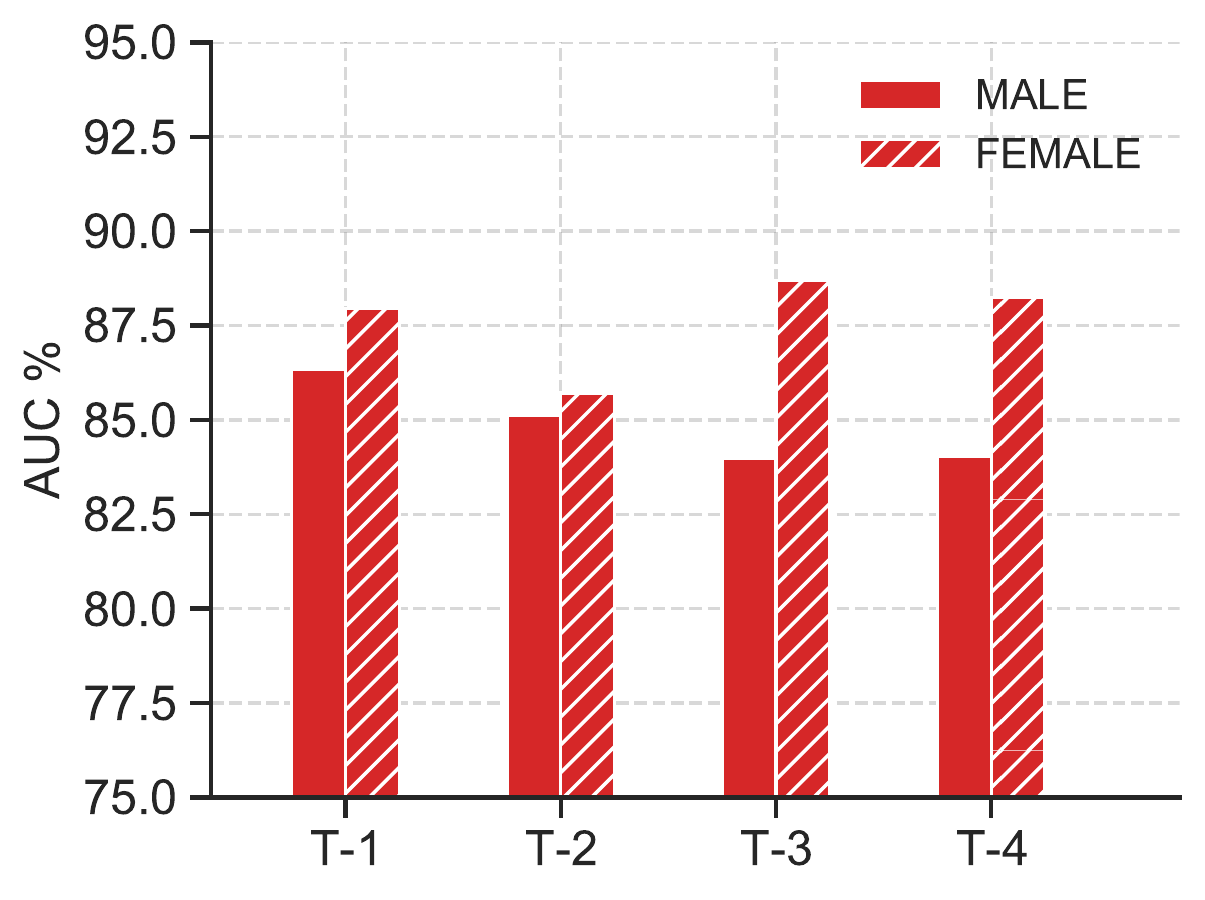}};

\node[inner sep=0pt,anchor=center] (main_fig) at (13,10) 
{\includegraphics[width=5cm,height=4cm]{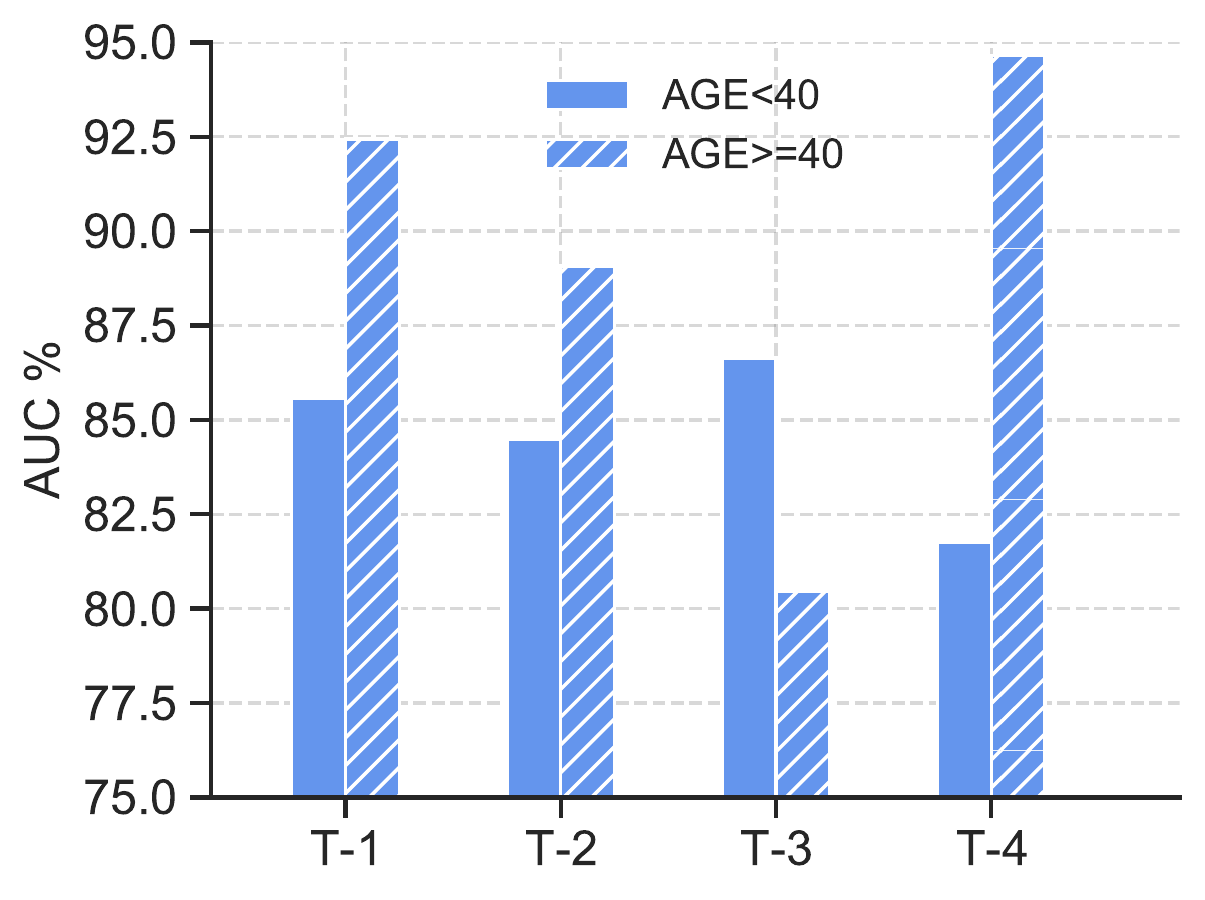}};

\node[font=\fontsize{8}{6}\selectfont,rotate=0,anchor=center] at (2,12.5) (l1) {Full Test set};
\node[font=\fontsize{8}{6}\selectfont,rotate=0,anchor=center] at (8,12.5) (l1) {Gender-split Test set};
\node[font=\fontsize{8}{6}\selectfont,rotate=0,anchor=center] at (13.5,12.5) (l1) {Age-split Test set};


\end{tikzpicture}

%% file: table_auc_sensitivity.tex
\begin{table*}[t]
\centering
\begin{tabular}{@{}ccccccccccccc@{}}
\toprule
\textbf{\begin{tabular}[c]{@{}c@{}}Performance\\ Measures\end{tabular}} & \textbf{} & \textbf{Team T-1} & \textbf{} & \textbf{Team T-2} & \textbf{} & \textbf{Team T-3} & \textbf{} & \textbf{Team T-4} &  & \textbf{Fusion} &  & \textbf{Baseline} \\ \midrule
AUC \% &  & \textbf{87.07} &  & 85.43 &  & 85.35 &  & 85.21 &  & 95.07 &  & 69.85 \\
\begin{tabular}[c]{@{}c@{}}Sensitivity\\ (at 95\% Specificity)\end{tabular} &  & 46.34 &  & 39.02 &  & \textbf{60.97} &  & 29.27 &  & 70.73 &  & 17.07 \\ \bottomrule
\end{tabular}
\caption{A comparison of AUC and sensitivity of top four teams, their score fusion and the baseline system.}
\label{table:auc_sensitivity}
\end{table*}